\documentclass[aps,prl,twocolumn,groupedaddress]{revtex4}
\usepackage{amssymb}
\usepackage{graphicx}

\begin{document}

\title{\bf Query complexity for searching multiple marked states from an unsorted database}


\author{Bin Shang}
\email[]{binshang@bit.edu.cn}
\affiliation{School of Computer Science $\&$ Technology,Beijing
Institute of Technology, No. 5, Zhongguancun Nandajie, Haidian
District, Beijing 100081, P. R. China}


\date{\today}
\begin{abstract}
An important and usual problem is to search all states we want from
a database with a large number of states. In such, recall is vital.
Grover's original quantum search algorithm has been generalized to
the case of multiple solutions, but no one has calculated the query
complexity in this case. We will use a generalized algorithm with
higher precision to solve such a search problem that we should find
all marked states and show that the practical query complexity
increases with the number of marked states. In the end we will
introduce an algorithm for the problem on a ``duality computer'' and
show its advantage over other algorithms.
\end{abstract}

\maketitle

\section{Introduction}
Since L. K. Grover \cite{1}\cite{2} discovered the quantum algorithm
for the unsorted database search problem with single marked state,
many improvements have been made on it. Among those, Michel Boyer et
al\cite{3}were the first to generalize it to the case of multiple
solutions; G. Brassard et al\cite{4}, P. H\o yer\cite{5}, G. L.
Long\cite{6} respectively improved Grover's algorithm and obtained
certainty in finding the single marked state in different ways.
Besides, many works\cite{3}\cite{7}\cite{8}\cite{9}\cite{10}
analyzed the query complexity and lower bounds of Grover's algorithm
or related algorithms for search problems. However, the query
complexity for searching all multiple marked states from an unsorted
database has not been addressed. This problem is very important when
recall is emphasized in searching multiple objects.
\section{Query complexity for searching all multiple marked states using generalized Long's algorithm}
\subsection{Generalized Long's algorithm}
A generalization of Long's algorithm\cite{6} for searching single
marked state with certainty in an unsorted database to the case of
multiple marked states can be easily shown like this: repeat Long's
algorithm until all marked states have been found.
\subsection{Problem description}
Let us define the problem as:\\
\emph{Under randomized conditions, search all $m$ marked states $
|\tau_1\rangle,|\tau_2\rangle,...,|\tau_m\rangle $from an unsorted
database with $N$ states
$|0\rangle,|1\rangle,\ldots,|\tau_1\rangle,|\tau_2\rangle,\ldots,|\tau_m\rangle,\ldots,|N-1\rangle$
with no less than overall probability of success $1-\delta$.}
\subsection{Solution and query complexity}
Now we analyze the number of queries needed to solve this problem.
Without loss of generality, we divide the process to find all $m$
marked states into $m$ steps. We devote $q_i$ as the queries needed
and $r_i$ as the times that we should run Long's algorithm, to find
$|\tau_i\rangle$; we devote $q$ as the total queries needed, and $r$
as the total times that we should run Long's algorithm. Besides, for
simplicity we suppose we could ``fortunately'' find every marked
state ultimately during corresponding queries in every step. Still,
we can educe the query complexity with high precision if $\delta$ is
small.
\subsubsection{Step 1}
Without loss of generality, we can find $|\tau_1\rangle$within
$q_1=\mathcal {O}(\sqrt\frac{N}{m})$ queries with certainty using
Long's algorithm. We should run the algorithm for only once, so here
$r_1=1$.
\subsubsection{Step 2}
We should stress that in this step we have probability $\frac{1}{m}$
to find $|\tau_1\rangle$ again in the first run of Long's algorithm;
if it occurs, we have to start the second run of the algorithm, and
thus we have probability $(\frac{1}{m})^2$ to find $|\tau_1\rangle$
again\ldots. As a result, to find $|\tau_2\rangle$ with probability
$1-\delta$,we should run Long's algorithm for $r_2$ times such that
\begin{displaymath}
\frac{m-1}{m}+\frac{1}{m}\times\frac{m-1}{m}+(\frac{1}{m})^2\times\frac{m-1}{m}+
\end{displaymath}
\begin{displaymath}
\ldots+(\frac{1}{m})^{r_2}\times\frac{m-1}{m}=1-\delta
\end{displaymath}
which equals to
\begin{displaymath}
(\frac{1}{m})^{r_2}=\delta
\end{displaymath}
Thus, we can get
\begin{displaymath}
r_2=\frac{\ln \delta^{-1}}{\ln{m}}
\end{displaymath}
and
\begin{displaymath}
q_2=\frac{\ln \delta^{-1}}{\ln m}\mathcal {O}(\sqrt\frac{N}{m})
\end{displaymath}
\subsubsection{\ldots\ldots}
\subsubsection{Step i}
In this step we have probability $\frac{i-1}{m}$ to find what we
have found in former
steps---$|\tau_1\rangle$,$|\tau_2\rangle$,\ldots,$|\tau_i-1\rangle$---again
in the first run of Long's algorithm; if it occurs, we have to start
the second run of the algorithm, and thus we have probability
$(\frac{i-1}{m})^2$ to find what we have found in former steps
again\ldots. As a result, to find $|\tau_i\rangle$ with probability
$1-\delta$,we should run Long's algorithm for $r_i$ times such that
\begin{displaymath}
\frac{m-i+1}{m}+\frac{1}{m}\times\frac{m-i+1}{m}+(\frac{1}{m})^2\times\frac{m-i+1}{m}+
\end{displaymath}
\begin{displaymath}
\ldots+(\frac{1}{m})^{r_i}\times\frac{m-i+1}{m}=1-\delta
\end{displaymath}
which equals to
\begin{displaymath}
(\frac{i-1}{m})^{r_i}=\delta
\end{displaymath}
Thus, we can get
\begin{displaymath}
r_i=\frac{\ln \delta^{-1}}{\ln{m}-\ln{(i-1)}}
\end{displaymath}
and
\begin{displaymath}
q_i=\frac{\ln \delta^{-1}}{\ln{m}-\ln{(i-1)}}\mathcal
{O}(\sqrt\frac{N}{m})
\end{displaymath}
\subsubsection{\ldots\ldots}
\subsubsection{Step m}
In this step we have probability $\frac{m-1}{m}$ to find what we
have found in former steps again in the first run of Long's
algorithm; if it occurs, we have to start the second run of the
algorithm, and thus we have probability $(\frac{m-1}{m})^2$ to find
what we have found in former steps again\ldots. As a result, to find
$|\tau_m\rangle$ with probability $1-\delta$,we should run Long's
algorithm for $r_m$ times such that
\begin{displaymath}
\frac{1}{m}+\frac{m-1}{m}\times\frac{1}{m}+(\frac{m-1}{m})^2\times\frac{1}{m}+
\end{displaymath}
\begin{displaymath}
\ldots+(\frac{m-1}{m})^{r_m}\times\frac{1}{m}=1-\delta
\end{displaymath}
which equals to
\begin{displaymath}
(\frac{m-1}{m})^{r_m}=\delta
\end{displaymath}
Thus, we can get
\begin{displaymath}
r_m=\frac{\ln \delta^{-1}}{\ln{m}-\ln{(m-1)}}
\end{displaymath}
and
\begin{displaymath}
q_m=\frac{\ln \delta^{-1}}{\ln{m}-\ln{(m-1)}}\mathcal
{O}(\sqrt\frac{N}{m})
\end{displaymath}
\subsubsection{Result}
To add up, we can obtain
\begin{eqnarray}
r=1+\sum_{k=1}^{m-1}\ln{\delta^{-1}}\ln^{-1}{(\frac{m}{k})}\\
q=[1+\sum_{k=1}^{m-1}\ln{\delta^{-1}}\ln^{-1}{(\frac{m}{k})}]\mathcal
{O}(\sqrt\frac{N}{m})
\end{eqnarray}
\subsection{Examples}
Let us see two examples of the result of $r$.
\subsubsection{Example 1}
\begin{figure}
\begin{center}
  \includegraphics[width=9cm,height=8cm]{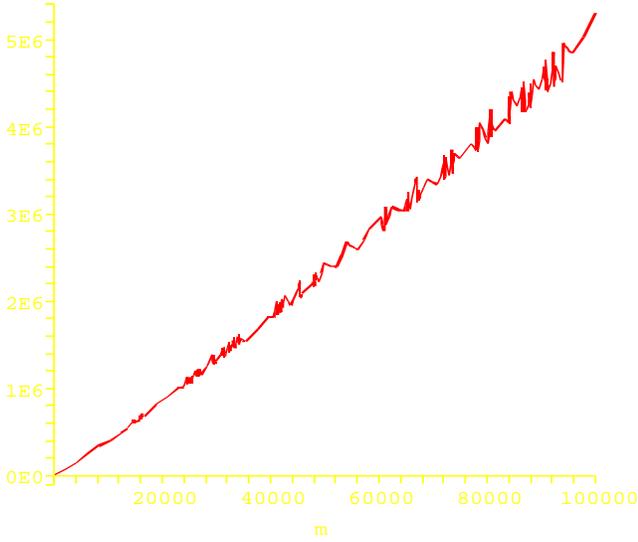}
  \caption{$f(m)$ when $\delta=0.01, 1\leq m\leq 100000$}\label{f1}
\end{center}
\end{figure}

\begin{figure}
\begin{center}
  \includegraphics[width=9cm,height=8cm]{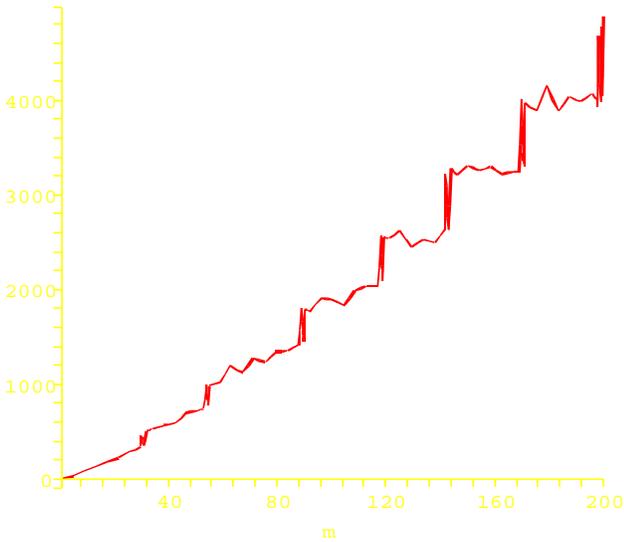}
  \caption{$f(m)$ when $\delta=0.01, 1\leq m\leq 200$}\label{f2}
\end{center}
\end{figure}

We set $\delta=0.01$, and devote
\begin{displaymath}
f(m)=1+\sum_{k=1}^{m-1}\ln{0.01^{-1}}\ln^{-1}{(\frac{m}{k})}
\end{displaymath}
$f(m)$ is plotted in FIG.\ref{f1} and FIG.\ref{f2}.
\subsubsection{Example 2}
\begin{figure}
\begin{center}
  \includegraphics[width=9cm,height=8cm]{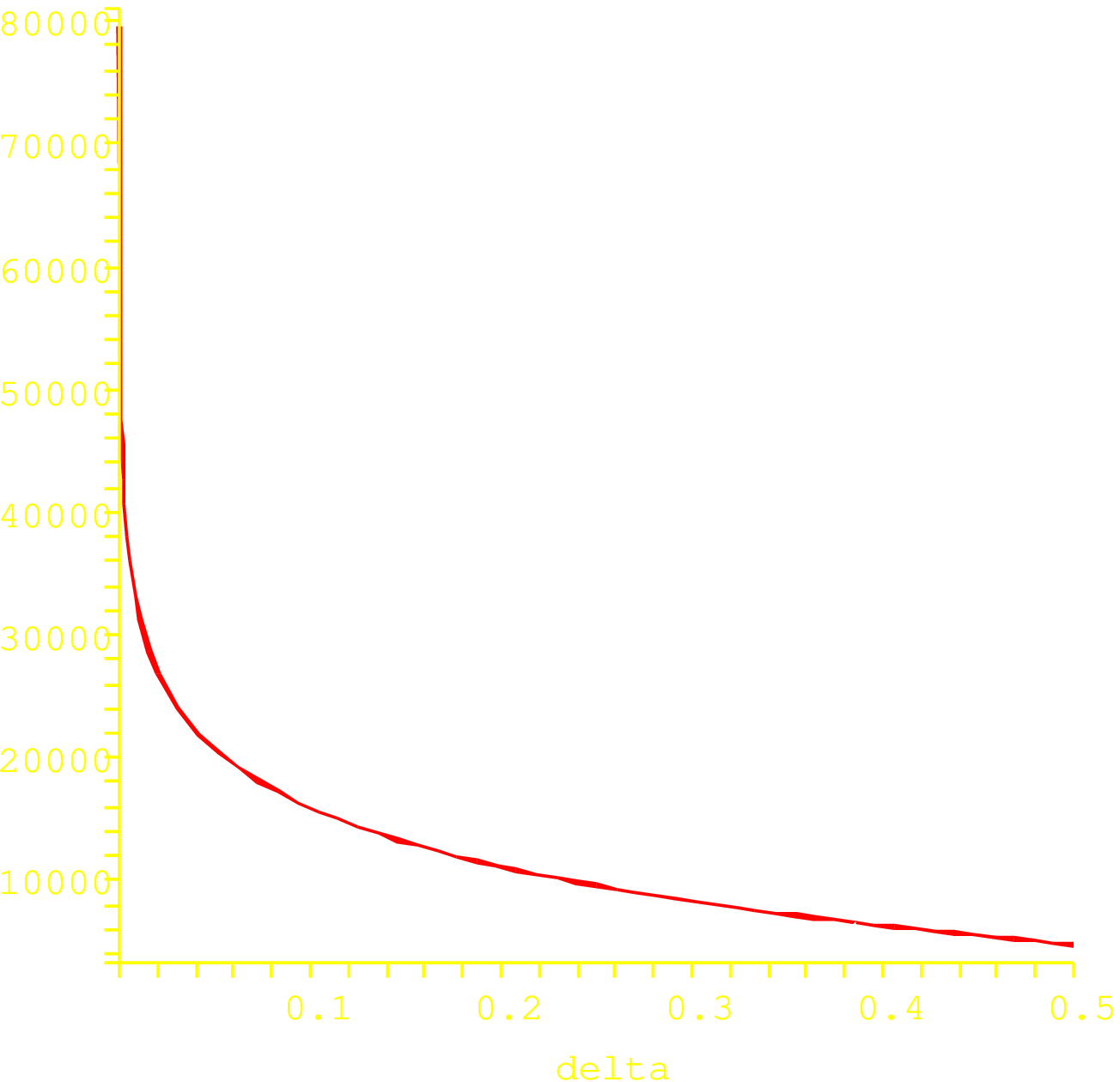}
  \caption{$f(\delta)$ when $m=1000, 0.00001\leq \delta\leq 0.5$}\label{f3}
\end{center}
\end{figure}

\begin{figure}
\begin{center}
  \includegraphics[width=9cm,height=8cm]{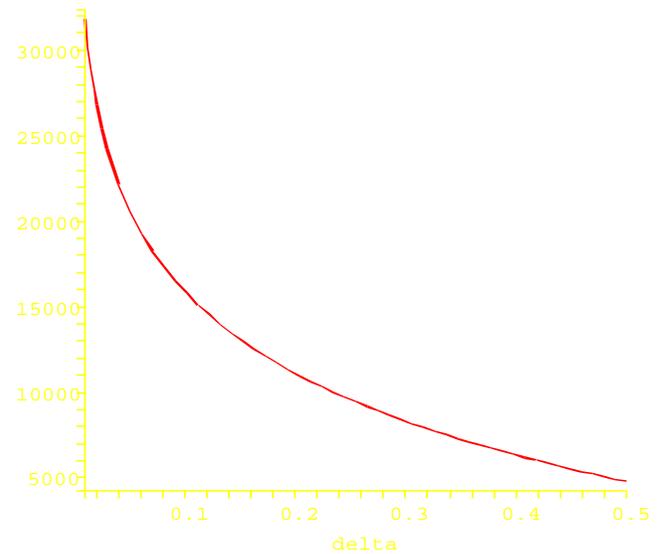}
  \caption{$f(\delta)$ when $m=1000, 0.01\leq \delta\leq 0.5$}\label{f4}
\end{center}
\end{figure}

We set $m=1000$, and devote
\begin{displaymath}
f(\delta)=1+\sum_{k=1}^{999}\ln{\delta^{-1}}\ln^{-1}{(\frac{1000}{k})}
\end{displaymath}
$f(\delta)$ is plotted in FIG.\ref{f3} and FIG.\ref{f4}.\\
From these figures, we can conclude that the query complexity will
increase rapidly with increase of the number of the marked states at
a specific probability of success, and will increase with precision
we need i.e. probability of success in finding every marked state.
\section{the search problem on a duality computer}
G. L. Long proposed a new quantum computing model---duality
computer\cite{11}---utilizing quantum system's wave-particle duality
which can achieve exponential speedup on unsorted database search
problems. Long proposed as well two search algorithms on a duality
computer respectively for the case of single marked state\cite{12}
and the case of multiple marked states. In the latter case, one can
find all $m$ marked states from an unsorted database with $N$ states
with certainty within $m\log{\frac{N}{m}}$ queries using $\log{N}$
``dubits''.\\
We should stress that in a duality computer, we can easily---within
time and space complexity $\mathcal {O}(1)$---delete the marked
states that we have found from the initial states. So we will not
meet such problems that with $m$ increasing the query complexity
increases as well when searching multiple marked states.
\section{Acknowledgements}
The author would like to thank Gui Lu Long for very helpful
discussion and thank Jia Qu Yi for his generous help in plotting the
function figures.
\begin{acknowledgments}
\end{acknowledgments}


\end{document}